\documentclass[aps,prx,preprint,superscriptaddress]{revtex4-1}

\usepackage[version=3]{mhchem}
\usepackage{bm}
\usepackage{amsmath}
\usepackage{ulem}
\usepackage{amssymb}
\usepackage{amsfonts}
\usepackage{euscript}
\usepackage{color}
\usepackage{epsfig}
\usepackage[section]{placeins}
\usepackage{gensymb}
\usepackage{array}
\usepackage{mathtools}

\makeatletter
\newcommand{\thickhline}{
    \noalign {\ifnum 0=`}\fi \hrule height 1pt
    \futurelet \reserved@a \@xhline
}
\newcolumntype{"}{@{\hskip\tabcolsep\vrule width 1pt\hskip\tabcolsep}}
\makeatother

\begin{document}

\title{Interplay of magnetic responses in all-dielectric oligomers to realize magnetic Fano resonances}

\author{Ben~Hopkins}
\email{ben.hopkins@anu.edu.au}
\affiliation{Nonlinear Physics Centre, Australian National University, Canberra, ACT 2601, Australia}

\author{Dmitry~S.~Filonov}
\affiliation{ITMO University, St. Petersburg 197101, Russia}

\author{Andrey~E.~Miroshnichenko}
\affiliation{Nonlinear Physics Centre, Australian National University, Canberra, ACT 2601, Australia}

\author{Francesco~Monticone}
\affiliation{Department of Electrical and Computer Engineering, The University of Texas at Austin, Austin, Texas 78712, USA}

\author{Andrea~Al\`u}
\affiliation{Department of Electrical and Computer Engineering, The University of Texas at Austin, Austin, Texas 78712, USA}

\author{Yuri~S.~Kivshar}
\affiliation{Nonlinear Physics Centre, Australian National University, Canberra, ACT 2601, Australia}
\affiliation{ITMO University, St. Petersburg 197101, Russia}




\begin{abstract}
We study the interplay between collective and individual optically-induced magnetic responses in quadrumers made of identical dielectric nanoparticles. Unlike their plasmonic counterparts, all-dielectric nanoparticle clusters are shown to exhibit multiple dimensions of resonant magnetic responses that can be employed for the realization of anomalous scattering signatures. We focus our analysis on symmetric quadrumers made from silicon nanoparticles and verify our theoretical results in proof-of-concept radio frequency experiments demonstrating the existence of a novel type of magnetic Fano resonance in nanophotonics.
\end{abstract}
\maketitle

A key concept in the study of optical metamaterials has been the use of geometry to engineer and boost the magnetic response from metallic nanostructures.~\cite{Smith2002, Shalaev2007, Liu2011}
Indeed, while simple metallic nanoparticles have a negligible magnetic response, a split ring resonantor, or a properly arranged cluster of three or more particles, is able to sustain localized magnetic resonances.~\cite{Pendry1999, Alu2006, Alu2009,Monticone2014}
An alternate route to optical magnetism is based on single nanoparticles made from high index dielectrics, because each such nanoparticle can exhibit an inherent magnetic response.~\cite{Evlyukhin2010, Garcia-Etxarri2011, Kuznetsov2012,  Evlyukhin2012NL}
In this work, we combine the use of the magnetic responses arising from {\it both}: (i) individual nanoparticles and (ii) design geometry, to provide {\it two} distinct dimensions for purely-magnetic dipolar response. 
Specifically, we present a comprehensive study of the magnetic interplay between individual and collective magnetic responses in symmetric clusters of four nanoparticles, known as {\it quadrumers}, and discuss the appearance of a novel type of {\it magnetic Fano resonance}.
The ability to achieve directional Fano resonances with magnetic-type responses was considered for particles with negative permeability~\cite{Lukyanchuk2010}, but here we demonstrate magnetic dipolar Fano resonances in the total cross section of nanostructures made of conventional materials.  
Nanoparticle quadrumers were chosen because they have previously been shown to exhibit significant magnetic responses when light couples into a resonant circulation of displacement current associated with a large magnetic dipole moment.~\cite{Brandl2006, Alu2006, Fan2010, Shafiei2013,Roller2015}
It was further demonstrated that, by breaking a quadrumer's geometric symmetry, it is possible to get interaction between this collective magnetic response and the in-plane electric response; an interaction that can lead to sharp magneto-electric hybrid Fano resonances.~\cite{Shafiei2013, Monticone2014}
However, as we show in the following, dielectric nanoparticles enable a different coupling mechanism between magnetic resonances, one which does not require breaking geometric symmetry; the $z$-polarized magnetic field\footnote{Where $z$ is the quadrumer's principle axis.} produced by a resonant circulation of current is, under certain conditions, able to couple to the inherent magnetic response of the individual nanoparticles. 
Here, we demonstrate that, by using an $s$-polarized plane wave at oblique incidence to induce a resonant circulating current across the cluster, we can couple the collective magnetic response of a symmetric all-dielectric quadrumer into the inherent magnetic response of its individual dielectric particles. 
The interference between magnetic responses can then be tailored to produce distinctive and sharp magnetic Fano resonances.
This form of optically-induced magnetic-magnetic coupling and interference is a unique characteristic of, properly designed, dielectric nanoclusters.
{
This work thereby presents a new way to tailor the magnetic responses of all-dielectric metamaterials and nanoantenna devices, a result that complements the recent interest in dielectric nanostructures that utilize simultaneous excitation and tailoring of electric and magnetic optical responses.~\cite{Moitra2013, Staude2013, Chong2014, Wu2014, Shcherbakov2014} 
}

\section{Results and Discussion}

\begin{figure}[!h]
\centering
\centerline{\includegraphics[width=0.5\textwidth]{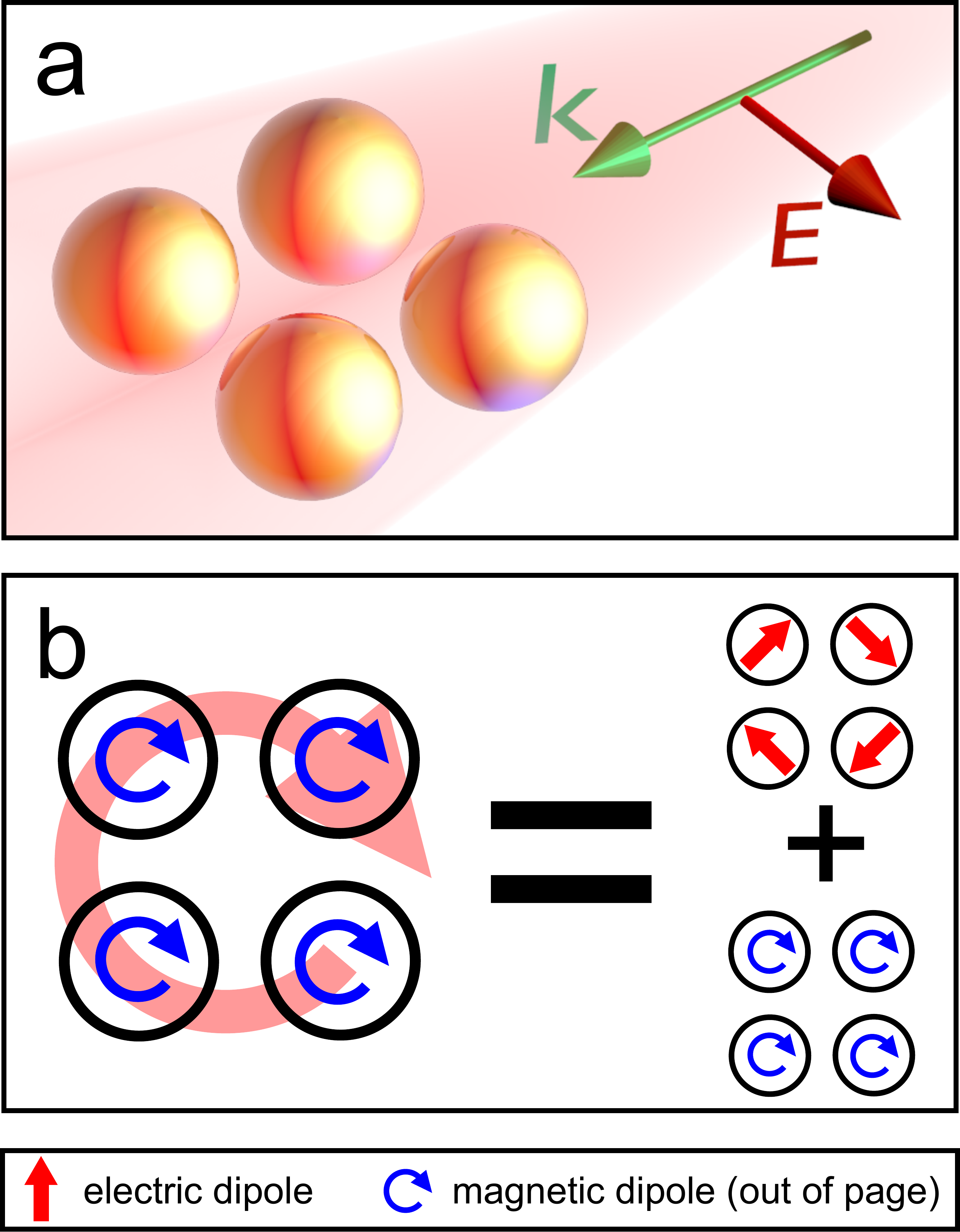}}
\caption{(a) Schematic of the problem: a symmetric nanosphere quadrumer is excited by an in-plane $s$-polarized plane wave {(both propagation and polarization vectors lie in the plane of the quadrumer)}.
(b) The combined magnetic response: a collective optically-induced magnetic resonance of the whole structure is accompanied 
by the magnetic responses of the individual particles.
}
\label{fig:schematic}
\end{figure}

{ Consider the optical response of a symmetric quadrumer when it is excited by a plane wave, whose propagation direction and electric field polarization lie in the plane of the quadrumer ($s$-polarization), as shown in Fig.~\ref{fig:schematic}a.  }
The conventional magnetic response to such an excitation is supported by a collective circulation of electric displacement current around all four nanoparticles.
However,  if the nanoparticles are made of a high-index dielectric material such as silicon, the individual nanoparticles are also able to couple directly with the applied magnetic field, sustaining internal circulating polarization currents. 
In this sense, a silicon quadrumer can produce magnetic responses to both electric {\it and} magnetic components of the applied plane wave, as is depicted in Fig.~\ref{fig:schematic}b.
{ In this figure, the red arrows denote an electric dipole, and the blue circular arrows denote the circulation of displacement current that supports an out-of-page magnetic dipole.
Importantly, } an interplay may be induced between the two magnetic responses through the locally-enhanced, $z$-polarized magnetic field sustained by the collective circulation of displacement current, which establishes a coupling channel to the magnetic responses of the individual nanoparticles.

\begin{figure}[!h]
\centering
\centerline{\includegraphics[width=0.9\textwidth]{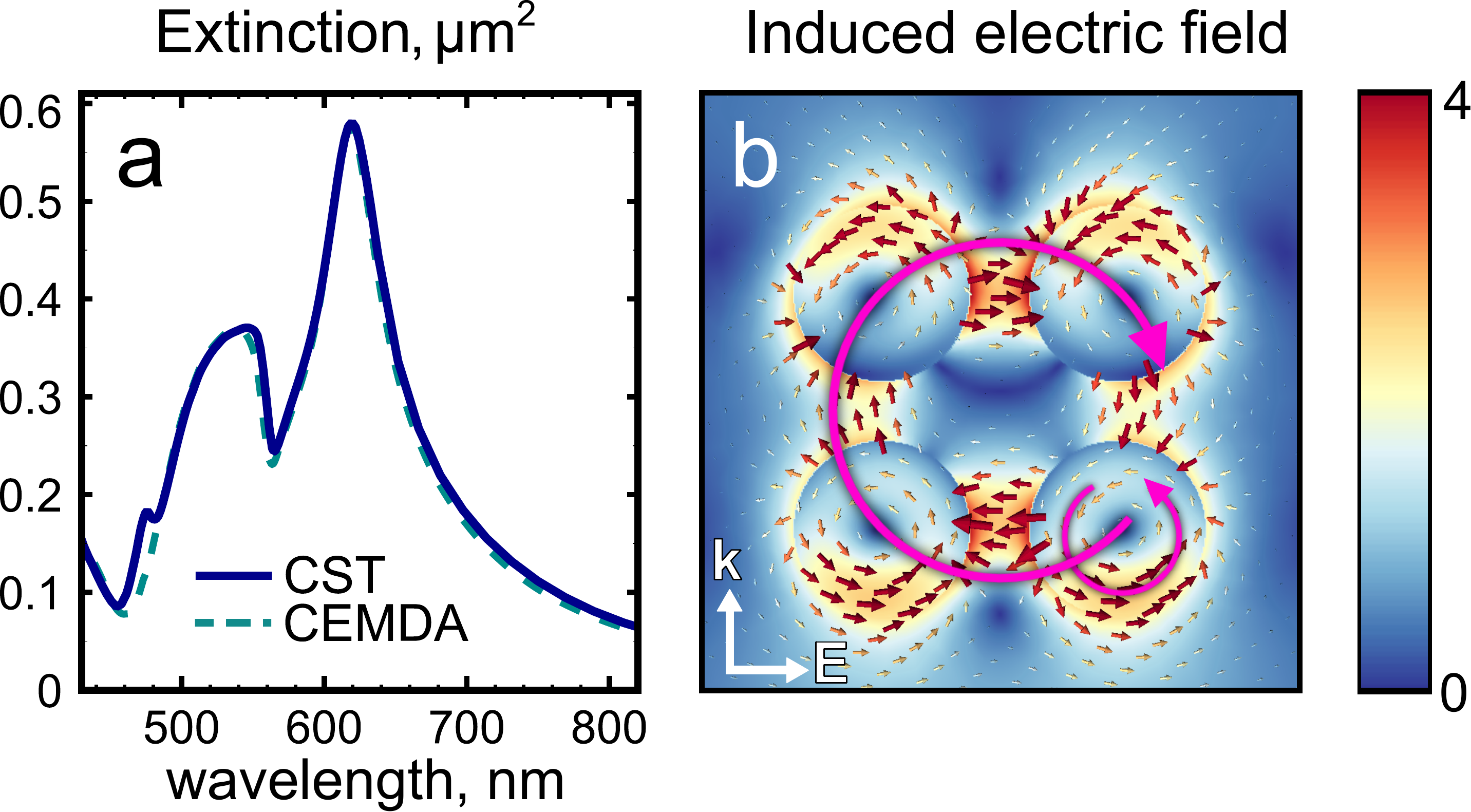}}
\caption{
(a) Extinction calculated using both the CEMDA and CST Microwave Studio for a quadrumer composed of $150\,\mathrm{nm}$ (diameter) silicon nanospheres, each separated by $50\,\mathrm{nm}$. (b) Associated near-field profile of the induced electric field at the silicon quadrumer's Fano resonance $(\lambda =570 \,\mathrm{nm})$; the superimposed cyan arrows indicate the direction of circulation for the electric field both external and internal to the individual spheres. Electric field amplitude is normalized to the amplitude of the incident {  plane wave, whose propagation direction ($\mathbf{k}$) and polarization ($\mathbf{E}$) are both parallel to the plane of the quadrumer}.  }
\label{fig:CSTvsCEMDA}
\end{figure}

We begin by understanding this magnetic interplay, for which we will model the interactions between nanoparticles using the {\it Coupled Electric and Magnetic Dipole Approximation}~\cite{Mulholland1994} (CEMDA).
The mathematical description of this dipole model in free space is based on the following equations:
\begin{subequations} \label{eq:dipole equations}
\begin{align}
\mathbf{p}_{i}  = \alpha_{E}\epsilon_{0}\mathbf{E_{0}}(\mathbf{r}_{i})&
+\alpha_{E}k^{2}
  \left(\underset{j\neq i}{{\sum}}\hat{G}_{0}(\mathbf{r}_{i},\mathbf{r}_{j}) \mathbf{p}_{j}
-\frac{1}{c_0}\,\nabla\times\hat{G}_{0}(\mathbf{r}_{i},\mathbf{r}_{j}) \cdot \mathbf{m}_{j}\right) ,\\
 \mathbf{m}_{i} =  \alpha_{H}\mathbf{H_{0}}(\mathbf{r}_{i})&
 +\alpha_{H}k^{2}  \left(\underset{j\neq i}{{\sum}}\hat{G}_{0}(\mathbf{r}_{i},\mathbf{r}_{j}) \mathbf{m}_{j}
+c_0\,\nabla\times\hat{G}_{0}(\mathbf{r}_{i},\mathbf{r}_{j}) \cdot \mathbf{p}_{j}\right),
\end{align}
\end{subequations}
where $\mathbf{p}_i$ ($\mathbf{m}_i$) is the electric (magnetic) dipole moment of the $i^{\mathrm{th}}$ particle, $\hat{G}_{0}(\mathbf{r}_{i},\mathbf{r}_{j}) $ is the free space dyadic Greens function between the $i^{\mathrm{th}}$ and $j^{\mathrm{th}}$ dipole, $\alpha_{E}$ ($\alpha_{H}$) is the electric (magnetic) polarizability of a particle, $c_0$ is the speed of light and $k$ is the free-space wavenumber. 
In Fig.~\ref{fig:CSTvsCEMDA}a, we show that this dipole model is accurate in modeling full-wave simulations of the silicon nanosphere quadrumers performed using CST Microwave Studio, confirming that, in this spectral range, the optical response of such nanoparticle quadrumers is dominated by dipole interactions among the individual particles.
{ This is the case because higher order coupling in such systems will occur only with smaller interparticle separations~\cite{Albella2013} (see also the Supporting Information).} 
Fig.~\ref{fig:CSTvsCEMDA}b  shows the electric field distribution at the collective magnetic resonance of the cluster, confirming that the response of the silicon quadrumer includes a collective circulation of electric field around the particles. 
However, we can also see the additional dynamic produced by dielectric nanoparticles: the magnetic response induced by circulating transverse electric dipoles is accompanied by a magnetic response from each of the individual nanoparticles.
This is the unique effect we are interested in.

\begin{figure}[!h]
\centering
\centerline{\includegraphics[width=0.75\textwidth]{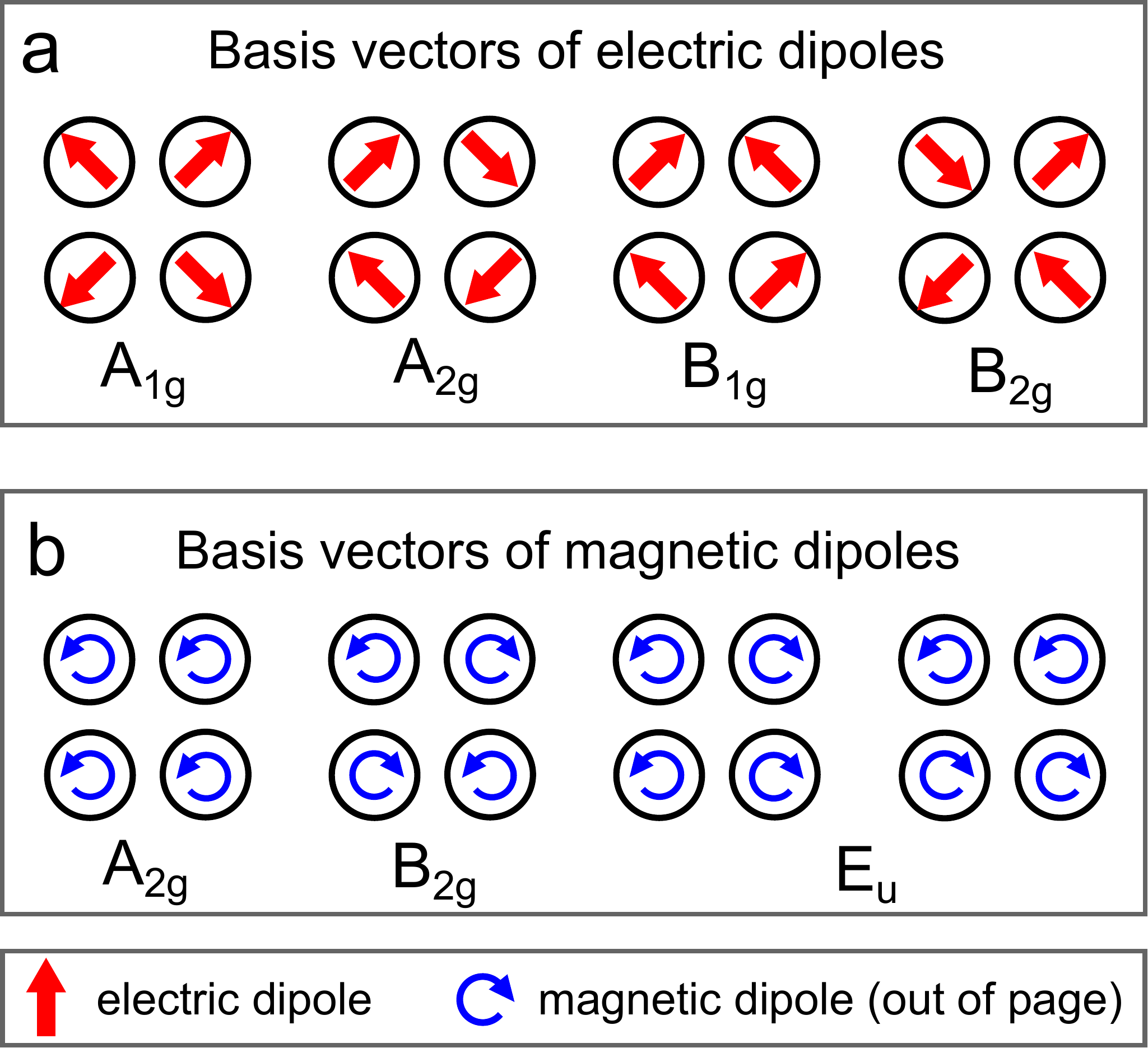}}
\caption{Basis vectors for the optical response of the quadrumer's electric and magnetic dipoles, named according to their associated irreducible representation.    Due to finite dimensions and symmetry constraints, each basis vector shown here is an eigenmode of the electric or magnetic dipole equation (Eq.~\ref{eq:decoupled equations}a or Eq.~\ref{eq:decoupled equations}b).   }
\label{fig:eigenmodes}
\end{figure}

The magnetic response in an individual silicon nanosphere is an internal circulation of polarization current, which can be seen in Fig.~\ref{fig:CSTvsCEMDA}b. 
In the CEMDA we choose to homogenize this circulating current distribution into a discrete source of magnetization current; the magnetic dipole.
This implies that the key to strong interaction between individual and collective magnetic responses resides in the strong coupling between induced electric and magnetic dipoles in each inclusion.
To describe this coupling, we can consider the eigenmodes of the quadrumer system.
As we will show, the eigenmodes of the full, electromagnetic, system can be constructed from the eigenmodes of the decoupled electric and magnetic dipole systems.
Therefore, we are going to break apart our dipole system of Eq.~\ref{eq:dipole equations} into two decoupled equations: one equation for the electric dipoles and one equation for the magnetic dipoles 
\begin{subequations} \label{eq:decoupled equations}
\begin{align}
\mathbf{p}_{i}  = \alpha_{E}\epsilon_{0}\mathbf{E_{0}}(\mathbf{r}_{i})&
+\alpha_{E}k^{2}
  \underset{j\neq i}{{\sum}}\hat{G}_{0}(\mathbf{r}_{i},\mathbf{r}_{j}) \mathbf{p}_{j}  \;,\\
 \mathbf{m}_{i} =  \alpha_{H}\mathbf{H_{0}}(\mathbf{r}_{i})&
+\alpha_{H}k^{2}
\underset{j\neq i}{{\sum}}\hat{G}_{0}(\mathbf{r}_{i},\mathbf{r}_{j}) \mathbf{m}_{j} \;.
\end{align}
\end{subequations}
We can then define two sets of eigenmodes from these equations: one for the electric dipoles and one for the magnetic dipoles.
Using state notation, we can refer to the electric dipole eigenmodes as $\left | \mathbf{e} \right \rangle$  and the magnetic dipole eigenmodes as $\left | \mathbf{h} \right \rangle$.
The eigenmode approach in the dipole approximation offers a huge simplification when combined with symmetry, because it allows us to determine certain eigenmodes without calculation.
The key is that, for both electric and magnetic dipole systems, the dipole approximation restricts the number of eigenmodes for each equation to twelve (three dimensions by four dipoles).
Each of these eigenmodes can only transform according to a single irreducible representation of the quadrumer's symmetry group,\cite{Dresselhaus2008} referred to as $\mathrm{D_{4h}}$.  
Further details regarding the $\mathrm{D_{4h}}$ symmetry group's irreducible representations and the implications for eigenmodes are provided in the { Supporting Information}. 
For the analysis here, it suffices to say that there are only eight irreducible representations in $\mathrm{D_{4h}}$ for twelve eigenmodes, and therefore the eigenspace associated with certain irreducible representations must be one-dimensional.  
Moreover, given that any vector in a one-dimensional space is an eigenvector by default, we are able to derive a number of eigenmodes by simply finding dipole moment profiles that transform according to certain irreducible representations.   
Eight such dipole moment profiles are shown in Fig.~\ref{fig:eigenmodes}.
Each of these dipole moment profiles is the sole basis vector for a single irreducible representation and is therefore an eigenmode of the electric or magnetic dipole equations in Eq.~\ref{eq:decoupled equations}, irrespective of wavelength or the choice of material, size, or any parameter which conserves the symmetry of the quadrumer.
It is worth noting that this same procedure for finding eigenmodes is applicable to other symmetries when using the dipole approximation, particularly the $\mathrm{D}_{n\mathrm{h}}$ symmetry groups for small values of $n$.  

\begin{figure}[!h]
\centering
\centerline{\includegraphics[width=0.8\textwidth]{{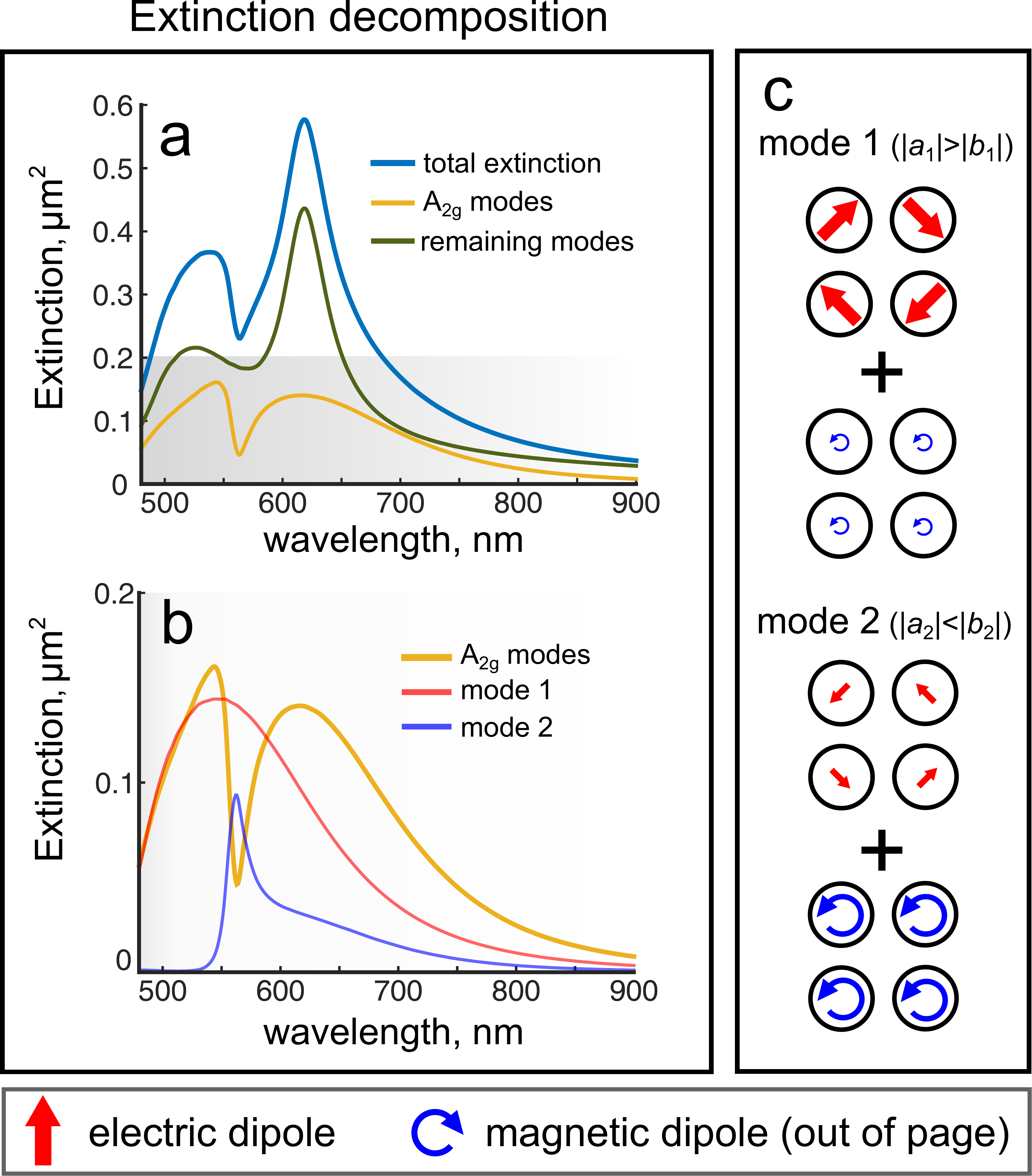}}}
\caption{Electromagnetic eigenmode decomposition of the silicon quadrumer, calculated using the CEMDA, showing (a) the contribution of the $\mathrm{A_{2g}}$ (magnetic-like) eigenmodes to the overall extinction.  In (b) we decompose the $\mathrm{A_{2g}}$ response  in terms of its two eigenmodes, showing the creation of a Fano resonance from  interference between them.  To support this we also show (c) the real components of the dipole moment profiles for each $\mathrm{A_{2g}}$ eigenmode, calculated at the wavelength of the Fano feature. }
\label{fig:interference}
\end{figure}

\begin{figure}[!t]
\centering
\centerline{\includegraphics[width=0.9\textwidth]{{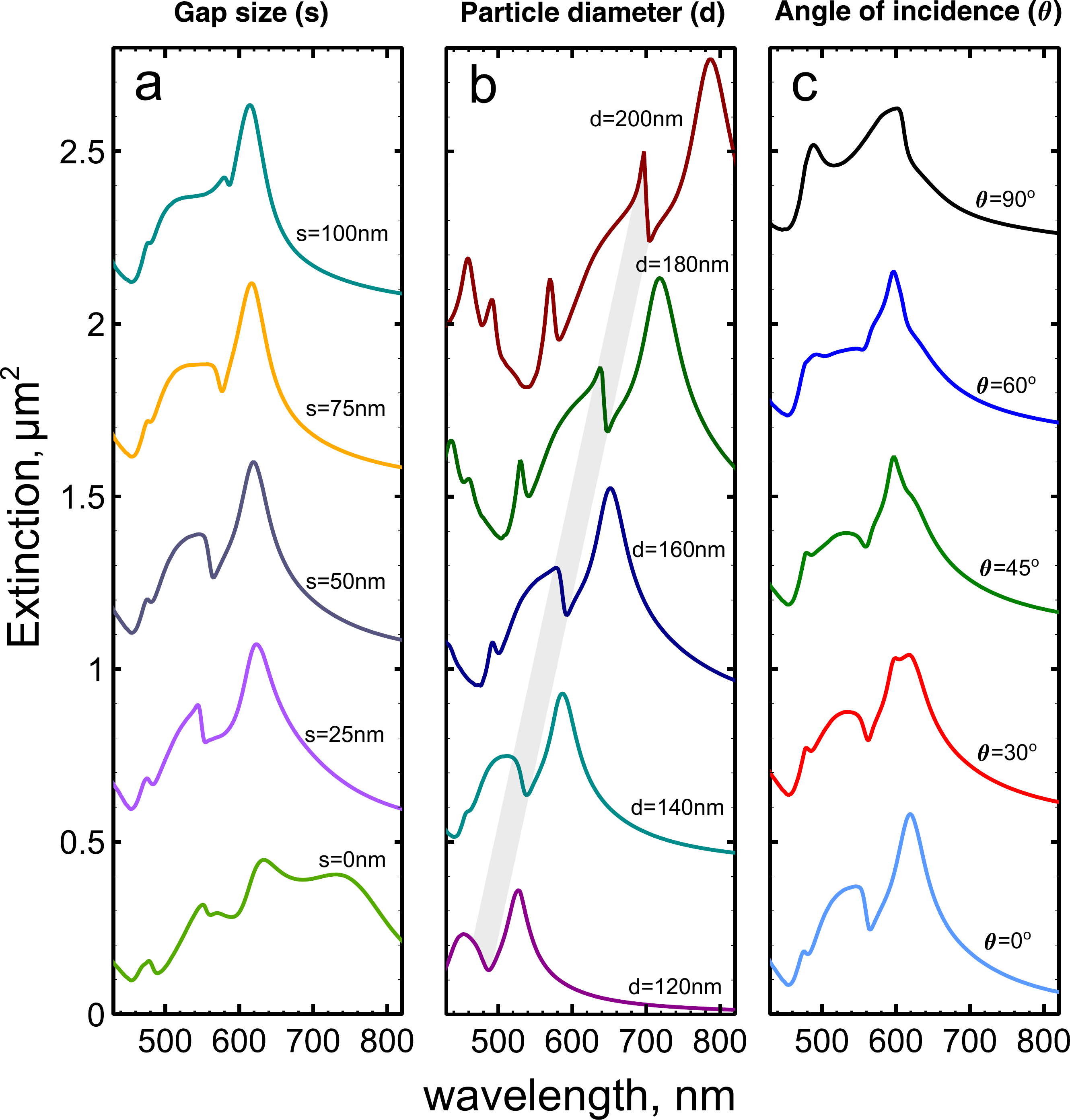}}}
\caption{Simulation results, from CST Microwave Studio, of the extinction caused by the silicon quadrumer in Fig.~\ref{fig:CSTvsCEMDA}, when changing (a) the size of the gap between nanoparticles, (b) the size of the individual nanoparticles and (c) the angle of incidence of the applied plane wave, with respect to the quadrumer plane (maintaining $s$-polarization).}
\label{fig:sweep}
\end{figure}

We need to now consider the interaction between electric and magnetic dipole systems.
An eigenmode of either electric or magnetic dipoles can be substituted into Eq.~\ref{eq:dipole equations} to determine the resulting state of magnetic or electric dipoles ($ | \mathbf{m_{(e)}} \rangle$ or $ | \mathbf{p_{(h)}} \rangle$, respectively) that it will induce due to magneto-electric coupling:
\begin{subequations}\label{eq:cross}
\begin{align}
\mathbf{m_{(e)}}_{i}  =\alpha_{H}  k^{2} {c_0} \underset{j\neq i}{{\sum}} \nabla\times\hat{G}_{0}(\mathbf{r}_{i},\mathbf{r}_{j}) \cdot \mathbf{e}_{j}\;,\\
\mathbf{p_{(h)}}_{i}  =\frac{-\alpha_{E}  k^{2}} {c_0} \underset{j\neq i}{{\sum}} \nabla\times\hat{G}_{0}(\mathbf{r}_{i},\mathbf{r}_{j}) \cdot \mathbf{h}_{j}\;.
\end{align}
\end{subequations}
However, it has been shown that the operation describing the coupling between states of electric and magnetic dipole moments, must commute with the geometry's symmetry operations.~\cite{HopkinsLiu2013}
Subsequently the $\left | \mathbf{e}\right \rangle$ and $ | \mathbf{m_{(e)}} \rangle$ dipole moments must both transform according to the same irreducible representation, and similarly for the $\left | \mathbf{h}\right \rangle$ and $ | \mathbf{p_{(h)}} \rangle$ dipole moments.
Therefore, if we consider the eigenmodes of electric and magnetic dipoles in Fig.~\ref{fig:eigenmodes}, the requirement for symmetry conservation specifies which electric dipole eigenmode can couple into which (if any) magnetic dipole eigenmode, and {\it vice versa}.
As a result, the  $\mathrm{A_{2g}}$ or $\mathrm{B_{1g}}$ eigenmodes ({\it cf.} Fig.~\ref{fig:eigenmodes}) are only able to magneto-electrically couple into the other $\mathrm{A_{2g}}$ or $\mathrm{B_{1g}}$ eigenmode, whereas the $\mathrm{A_{1g}}$ and $\mathrm{B_{2g}}$ eigenmodes are not able to couple into any eigenmodes due to a symmetry mismatch.
This conclusion allows us to take our analysis of the decoupled electric and magnetic dipole equations, and use it to determine the full, {\it electromagnetic}, eigenmodes of the real system.
Indeed, we would not necessarily expect an eigenmode of this complex scattering system to be a current distribution described by purely electric or magnetic dipoles; we would expect it to be a combination of both.
If we now restrict ourselves to considering only the $\mathrm{A_{2g}}$ (magnetic-like) eigenmodes of an all-dielectric quadrumer, the two eigenmodes from the decoupled electric and magnetic dipole equations ({\it cf.} Fig.~\ref{fig:eigenmodes}) form basis vectors for the $\mathrm{A_{2g}}$ eigenmodes, $| \mathbf{v}_x \rangle$, of the electromagnetic system.
\begin{align}
| \mathbf{v}_x \rangle = a_x  \left ( \!\!
\begin{array}{c}
| \mathbf{e} \rangle \\
0
\end{array}
\!\!\right) +
b_x \left (\!\! 
\begin{array}{c}
0 \\
| \mathbf{h} \rangle
\end{array}
\!\! \right)\,,\label{eq:EMeigenmode}
\end{align}
where $a_x$ and $b_x$ are complex scalars.  
Notably, we should expect to have two distinct electromagnetic eigenmodes here because there are two distinct basis vectors and, subsequently, a two-dimensional eigenspace for the quadrumer's response.
To derive expressions for $a_x$ and $b_x$, we can write the electromagnetic eigenvalue equation for  Eq.~\ref{eq:dipole equations}:
\begin{subequations} \label{eq:eigenmode equations}
\begin{align}
a_x \mathbf{e}_{i}  &= a_x \lambda_x  \alpha_{E}\epsilon_{0}\mathbf{e}_i+\alpha_{E}k^{2}
  \left(\underset{j\neq i}{{\sum}}a_x\hat{G}_{0}(\mathbf{r}_{i},\mathbf{r}_{j}) \mathbf{e}_{j}
- \frac{b_x}{{c_0}} \,\nabla\times\hat{G}_{0}(\mathbf{r}_{i},\mathbf{r}_{j}) \mathbf{h}_{j}\right) ,\\
b_x \mathbf{h}_{i} &= b_x  \lambda_x  \alpha_{H}\mathbf{h}_{i}
+\alpha_{H}k^{2}  \left(\underset{j\neq i}{{\sum}}b_x\hat{G}_{0}(\mathbf{r}_{i},\mathbf{r}_{j}) \mathbf{h}_{j}
+a_x {c_0}\, \nabla\times\hat{G}_{0}(\mathbf{r}_{i},\mathbf{r}_{j}) \mathbf{e}_{j}\right),
\end{align}
\end{subequations}
where $\lambda_x$ is the eigenvalue of the electromagnetic eigenmode.
However, Eqs.~\ref{eq:eigenmode equations}a and  \ref{eq:eigenmode equations}b are not independent\footnote{
Further details of this relation are provided in the { Supporting Information}.
}
 because one can be obtained from the other using duality transformations.~\cite{Jackson}
In other words, a solution for $a_x$ and $b_x$, that satisfies {\it either} Eq.~\ref{eq:eigenmode equations}a or Eq.~\ref{eq:eigenmode equations}b, must also satisfy the complete Eq.~\ref{eq:eigenmode equations} and describe an electromagnetic eigenmode of Eq.~\ref{eq:EMeigenmode}.
So, if the basis vectors  $\left | \mathbf{e}\right \rangle$ and $\left | \mathbf{h} \right \rangle $ are normalized to a magnitude of one, we can project Eq.~\ref{eq:eigenmode equations}a and Eq.~\ref{eq:eigenmode equations}b onto $ \mathbf{e}_{i}$ and $ \mathbf{h}_{i}$ (respectively) and sum over all $i$, to obtain the two solutions for Eq.~\ref{eq:eigenmode equations}:
\begin{align}
\frac{a_1}{b_1} &=
\frac{  \underset{i}{\sum} \, \mathbf{e}_i^{*}\cdot \mathbf{p_{(h)}}_{i} }
{\alpha_{\scriptscriptstyle E} \epsilon_0 (\lambda_{ \mathbf{e} } -  \lambda_1)}
 \;,  \label{eq:result1} \\[1 ex]
\frac{b_2}{a_2} &=
 \frac{\underset{i}{\sum} \mathbf{h}_i^{*}\cdot \mathbf{m_{(e)}}_{i}  }
{\alpha_{\scriptscriptstyle  H}(\lambda_{ \mathbf{h}} -  \lambda_2)}
 \;,  \label{eq:result2}
\end{align}
where $\lambda_{ \mathbf{e} }$ and $\lambda_{ \mathbf{h}} $ are the eigenvalues of $| \mathbf{e} \rangle$ and $| \mathbf{h} \rangle$  in the decoupled electric and magnetic dipole equations in Eq.~\ref{eq:decoupled equations}:
\begin{subequations} \label{eq:decoupled eigenequation}
\begin{align}
\alpha_{E}\epsilon_{0}\lambda_{ \mathbf{e}}  &
= 1 -\alpha_{E}k^{2}
  \underset{i,\,j\neq i}{{\sum}}\mathbf{e}_i^{*}\cdot \hat{G}_{0}(\mathbf{r}_{i},\mathbf{r}_{j}) \mathbf{e}_{j}  \;,\\
\alpha_{H}\lambda_{ \mathbf{h}}  &
= 1 -\alpha_{H}k^{2}
  \underset{i,\,j\neq i}{{\sum}}\mathbf{h}_i^{*}\cdot \hat{G}_{0}(\mathbf{r}_{i},\mathbf{r}_{j}) \mathbf{h}_{j}  \;.
\end{align}
\end{subequations}
The two ratios in Eq.~\ref{eq:result1} and Eq.~\ref{eq:result2} each describe a distinct eigenmode for the  electromagnetic system according to Eq.~\ref{eq:EMeigenmode}.
It is worth noting that, if either numerator or denominator go to zero in Eq.~\ref{eq:result1} or Eq.~\ref{eq:result2}, the starting basis vectors of the decoupled electric or magnetic dipole equations were already eigenmodes of the full system.
In regard to the $\mathrm{A_{2g}}$ (magnetic-like) eigenmodes of an all-dielectric quadrumer, our derivations show that, when both eigenmodes are far from resonance and magneto-electric coupling can be neglected, one eigenmode will be purely composed of electric dipoles and the other will be purely  composed of magnetic dipoles.
However, as they approach resonance, magneto-electric coupling cannot be neglected and the two ratios of $a_x$ and $b_x$ will need to be calculated to determine their form as electromagnetic eigenmodes (see Fig.~\ref{fig:interference}c).
In such a situation, the two eigenmodes, $| \mathbf{v}_1\rangle$ and $| \mathbf{v}_2\rangle$, are explicitly nonorthogonal to each other with the overlap defined by:
\begin{align}
 \langle \mathbf{v}_1|\mathbf{v}_2  \rangle =  a_1 ^* a_2 + b_1 ^* b_2\,. 
\label{eq:overlap}
\end{align}
We can therefore expect interference between these eigenmodes as they approach resonance.~\cite{HopkinsPoddubny2013}
 Indeed, as shown in Fig.~\ref{fig:interference}, the Fano resonance feature in the silicon quadrumer of Fig.~\ref{fig:CSTvsCEMDA} is produced entirely by the interference between these two eigenmodes.
To serve as a broader qualification of this magnetic Fano resonance feature, we also investigate its parameter dependence.
In Fig.~\ref{fig:sweep}a and \ref{fig:sweep}b, we vary the size of the gap between nanoparticles and the size of each individual nanoparticle, respectively.
{
It can be seen that the Fano resonance is dependent on a number of these coupling parameters, but in a way that is different to electric Fano resonances in plasmonic oligomers.  
Perhaps most notably, the magnetic Fano resonance is significantly shifted with small changes in size of the constituent nanoparticles, and closely spaced nanoparticles are favorable for electric Fano resonances in plasmonic oligomers, but can destroy the magnetic Fano resonance here.~\cite{Hentschel2010, RahmaniLukyanchuk2013}
More specifically, the magnetic Fano resonance feature is heavily dependent on the spacing between neighboring nanoparticles; increasing or decreasing the spacing can quickly diminish the Fano feature.
}
On the other hand, varying the size of the particles while holding the gap between particles constant is able to conserve the Fano resonance and shift it spectrally (shown in the shaded region of Fig.~\ref{fig:sweep}b).
This behavior can be expected because the action of increasing the size of the particles while holding the gap constant is very similar to uniformly scaling Maxwell's equations, and silicon permittivity has relatively minor dispersion in this spectral range.~\cite{Palik}   
Otherwise, the final parameter we consider in Fig.~\ref{fig:sweep}c, is the angle of incidence to address the practical limitations for in-plane excitation of nanoscale optical structures.
It can be seen that the Fano resonance will persist up to roughly 45\degree~incidence, beyond which the magnetic response is dominated by the normal-incidence responses.
\newpage

{\it Experimental verification.} To verify and validate the theoretical arguments presented above, we look for the existence of magnetic Fano resonances in a high index cluster experimentally. 
In this regard, one practical option is to mimic the scattering properties of silicon nanoparticles using MgO-$\mathrm{TiO_2}$ ceramic spheres characterized by dielectric constant of 16 and dielectric loss factor of $(1.12  -  1.17)\times10^{-4}$, measured at {$9 - 12$ GHz}. 
These ceramic spheres in the microwave range therefore have very similar properties to silicon nanospheres in the optical range and they are subsequently a useful macroscopic platform on which to prototype silicon nanostructures. 
Here, they allow us to perform a `proof of concept' investigation into the properties of an isolated quadrumer with much more signal than would be expected from a single silicon nanosphere quadrumer. Indeed, such spheres have been used previously to predict the behavior of silicon nanoantennas.{~\cite{Filonov2012, Geffrin2012, Filonov2014}}
\begin{figure}[!h]
\centering
\centerline{\includegraphics[width=0.9\textwidth]{{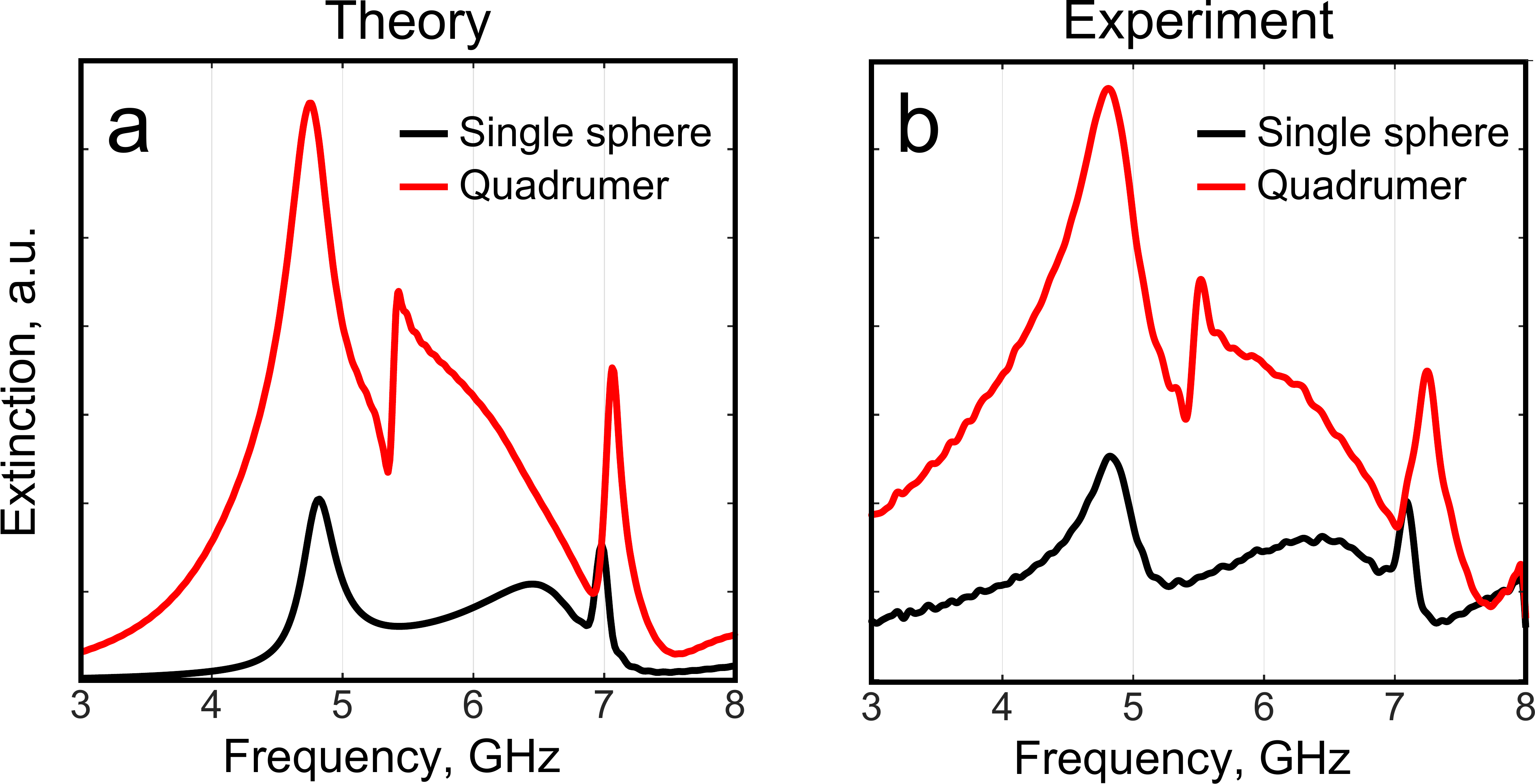}}}
\caption{The (a) CST simulation results and (b) experimental measurements of extinction for a quadrumer made of four MgO-$\mathrm{TiO_2}$ ceramic spheres.  We also show the extinction from a single MgO-$\mathrm{TiO_2}$ sphere for reference.  In both simulation and experiment we observe the existence of a sharp Fano resonance occurring at 5.4 GHz. }
\label{fig:experiment}
\end{figure}
The MgO-$\mathrm{TiO_2}$ quadrumer consists of  four dielectric spheres with diameter $d=15\mathrm{mm}$, and the size of the gap between the particles is $s=5\, \mathrm{ mm}$.
The experimentally measured, and numerically calculated, total scattering of the quadrumer  structure are shown in Fig.~\ref{fig:experiment}.
It can clearly be seen that a magnetic Fano resonance is produced at 5.4 GHz, in both simulation and experiment.
This is the first example of a magnetic-magnetic Fano resonance in a single symmetric metamolecule.
Notably, this Fano resonance occurs in a spectral range where the single particle is not at resonance, which demonstrates its collective nature. 
Indeed, it appears near the intersection of the single particle's electric and magnetic scattering contributions, reflecting that the overlap of eigenmodes (Eq.~\ref{eq:overlap}) is dependent on both electric and magnetic dipole polarizabilities.  

\section{Conclusions}

We have presented a comprehensive study of the interplay between the collective optically-induced magnetic responses of all-dielectric quadrumers and the individual magnetic responses of their constituent dielectric nanoparticles.
We have been able to establish the theoretical basis behind the interaction between collective and individual magnetic responses in all-dielectric structures, providing a quantitative prediction 
 of the interference between the quadrumer's magnetic responses leading to magnetic-magnetic Fano resonance features.
We have also been able to experimentally observe the existence of a sharp magnetic-magnetic Fano resonance in a dielectric quadrumer. 
Such Fano resonance features demonstrate the unique potential that suitably designed dielectric nanoclusters have for scattering engineering at the nanoscale, opening exciting and unexplored opportunities for dielectric nanophotonics.

\section{Methods}
To fasten together the MgO-$\mathrm{TiO_2}$ ceramic spheres for the experiment, we used a custom holder made of a styrofoam material with dielectric permittivity of 1 (in the microwave frequency range).
To approximate plane wave excitation, we employed a rectangular horn antenna (TRIM $0.75 - 18$ GHz ; DR) connected to the transmitting port of a vector network analyzer (Agilent E8362C).
The quadrumer was then located in the far-field of the antenna, at a distance of approximately 2.5 m, and a second horn antenna  (TRIM $0.75 - 18$ GHz ; DR) was used as a receiver to observe the transmission through the quadrumer.
The extinction measurement shown in Fig.~\ref{fig:experiment}b was then obtained as the difference between the measured transmission and unity transmission ({\it i.e.}, with no quadrumer).
For the theory, CST simulations were performed assuming plane wave excitation on a quadrumer located in free space. 
The CEMDA simulations, seen in Fig.~\ref{fig:CSTvsCEMDA}a, used electric and magnetic dipole polarizabilities that were derived from the scattering coefficients of Mie theory.~\cite{Mie1908}
In all simulations, silicon permittivity data was taken from Palik's Handbook~\cite{Palik} and the MgO-$\mathrm{TiO_2}$ spheres were assumed be dispersionless with a dielectric constant of 16 and dielectric loss factor of $(1.12  -  1.17)\times10^{-4}$.

{
\section{Acknowledgements}
This work was supported by the Australian Research Council.   F.M. and A.A. have been supported through the U.S. Air Force Office of Scientific Research with grant No. FA9550-13-1-0204, P02.   The measurements were supported by the Government of the Russian Federation (grant 074-U01), the Ministry of Education and Science of the Russian Federation, Russian Foundation for Basic Research, Dynasty Foundation (Russia).

}
\bibliographystyle{apsrev4-1}
\bibliography{bibliography}
\newpage
\appendix
\section{$\mathbf{D_{4h}}$ symmetry}

The eight irreducible representations of the $\mathrm{D_{4h}}$ symmetry group are depicted as the rows in Table~\ref{tab:D4h} and the columns correspond to symmetry operations, being: rotations ($\hat C$), reflections ($\hat \sigma$), inversions ($\hat i$), improper rotations ($\hat S$) and the identity ($\hat E$).
Each irreducible representation describes a distinct set of transformation behavior under the quadrumer's symmetry operations.
It follows that any given eigenmode can only transform according to a single irreducible representation and, consequently, that eigenmodes belonging to different irreducible representations must be orthogonal.
\begin{table}
\begin{tabular}{c | c  c   c   c  c  c  c  c  c  c  c}
~ & $\,\hat E\,$  & $\,2 \hat C_{4}\,$   & $\,\hat C_2\,$ & $\,2 \hat C'_2\,$ &  $\,2 \hat C''_2\,$ & $\;\hat i\;$  & $\,2 \hat S_4\,$ &  $\, \hat \sigma_h\,$  & $\,2 \hat \sigma_v\,$ & $\,2  \hat \sigma_d \,$ \\ [0.5ex]
\hline \\[-1ex]
$\mathrm{A_{1g}}\;$ & 1 &  1 &  1&  1  &  1 &  1 &  1 &  1 &  1 &  1   \\
$\mathrm{A_{2g}}\;$ & 1 &  1 &  1& -1  & -1 &  1 &  1 &  1 & -1 & -1   \\
$\mathrm{B_{1g}}\;$ & 1 & -1 &  1&  1  & -1 &  1 & -1 &  1 &  1 & -1   \\
$\mathrm{B_{2g}}\;$ & 1 & -1 &  1& -1  &  1 &  1 & -1 &  1 & -1 &  1   \\
$\mathrm{E_{u}}\;$   & 2 &  0 & -2&  0  &  0 &  2 &  0 & -2 &  0 &  0   \\
$\mathrm{A_{1u}}\;$ & 1 &  1 &  1&  1  &  1 & -1 &  1 & -1 & -1 & -1   \\
$\mathrm{A_{2u}}\;$ & 1 &  1 &  1& -1  & -1 & -1 &  1 & -1 &  1 &  1   \\
$\mathrm{B_{1u}}\;$ & 1 & -1 &  1&  1  & -1 & -1 & -1 & -1 & -1 &  1   \\
$\mathrm{B_{2u}\;}$ & 1 & -1 &  1& -1  &  1 & -1 & -1 & -1 &  1 & -1   \\
$\mathrm{E_{g}}\;$   & 2 &  0 & -2&  0  &  0 & -2 & -2 &  2 &  0 &  0   \\
 [1ex]
\hline
\end{tabular}\\~
\caption{Character table for the $\mathrm{D_{4h}}$ symmetry group.  The rows correspond to different irreducible representations and the columns are the symmetry operations.  Each number in the table is the character (trace) of the matrix representation of each symmetry operation for the given irreducible representation.}
\label{tab:D4h}
\end{table}
\newpage

\section{Duality transformations in dipole systems}
We can consider the electric and magnetic field radiated  by an arbitrary system of electric and magnetic dipoles:
\begin{align}
 \mathbf{E_{r}}(\mathbf{r}_i)  &= \frac{k^{2}}{\epsilon_0}
  \left(\underset{j\neq i}{{\sum}}\hat{G}_{0}(\mathbf{r}_{i},\mathbf{r}_{j}) \mathbf{p}_{j}   
- \frac{1}{{c_0}} \,\nabla\times\hat{G}_{0}(\mathbf{r}_{i},\mathbf{r}_{j}) \mathbf{m}_{j}\right) \;,\\
 \mathbf{H_{r}}(\mathbf{r}_i)  &=  k^{2} 
\left(\underset{j\neq i}{{\sum}}\hat{G}_{0}(\mathbf{r}_{i},\mathbf{r}_{j}) \mathbf{m}_{j} 
+ {c_0}\, \nabla\times\hat{G}_{0}(\mathbf{r}_{i},\mathbf{r}_{j}) \mathbf{p}_{j}\right)\;.
\end{align}
These two equations are, in fact, equivalent to each other under the duality transformation where $\mathbf{E_r} \rightarrow \sqrt{\frac{\mu_0}{\epsilon_0}} \mathbf{H_r}$, $\mathbf{H_r} \rightarrow -\sqrt{\frac{\epsilon_0} {\mu_0}} \mathbf{E_r}$, $\mathbf{p} \rightarrow \mathbf{m}/c_0$, $\mathbf{m} \rightarrow - c_0 \mathbf{p}$.
This follows directly from the duality transformation of the electrical displacement and $\mathbf{B}$-field, when expressed in terms of electric and magnetic fields, polarization, and magnetization.
Subsequently, if we return to the equations of the Coupled Electric and Magnetic Dipole Approximation (CEMDA) and rewrite them in terms of the radiated electric and magnetic field, we get:
\begin{align}
\mathbf{p}_{i}  &= \alpha_{E}\epsilon_{0}\mathbf{E_{0}}(\mathbf{r}_{i})
+\alpha_{E} \epsilon_{0} \mathbf{E_{r}}(\mathbf{r}_i) \;,\\
 \mathbf{m}_{i}  &= \alpha_{H}\mathbf{H_{0}}(\mathbf{r}_{i})
+\alpha_{H} \mathbf{H_{r}}(\mathbf{r}_i) \;.
\end{align}
Then, after dividing through by the respective dipole polarizabilities, it is clear to see that these equations simply state that the total field at any point is the sum of radiated and incident field.  
As such, the equivalence of radiated electric and magnetic fields under duality transformations is sufficient to make the two CEMDA equations equivalent under duality transformations.  

\newpage
\section{Near field accuracy of the dipole model}

\begin{figure}[!h]
\centerline{\includegraphics[width=0.9\textwidth]{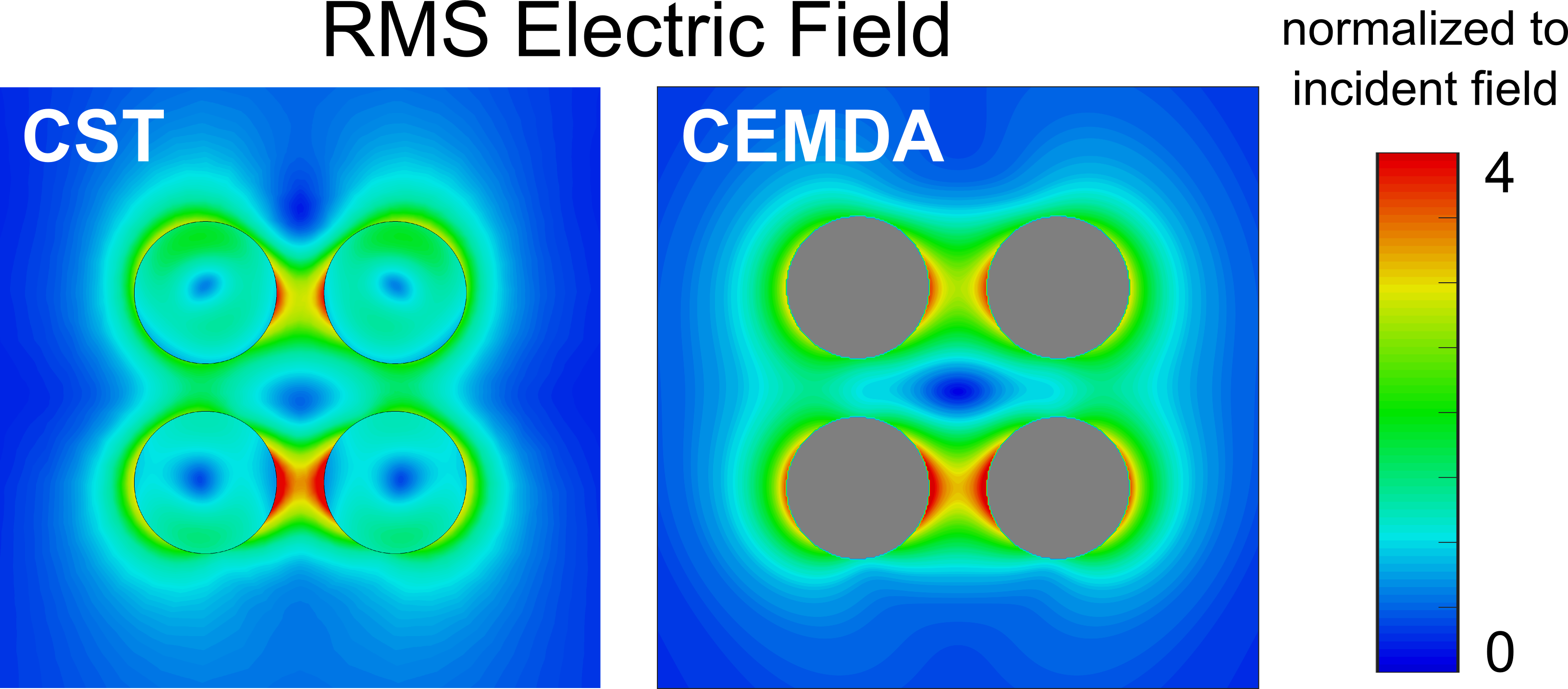}}
\caption{
The distribution of the Root Mean Square (RMS) scattered electric field amplitude of the silicon quadrumer used in Fig.~2 in the main text, calculated using CST and the Coupled Electric and Magnetic Dipole Aprroximation (CEMDA).  The simulations were performed at the wavelength of the quadrumer's Fano resonance (also used in Fig.~2b), and show that the CEMDA is accurate in predicting the near field behavior of the quadrumer we consider in the main text.   
}
\label{fig:comparison}
\end{figure}

The accuracy of the Coupled Electric and Magnetic Dipole Approximation (CEMDA) was demonstrated in Fig.~2a when modeling the extinction of the silicon nanoparticle quadrumer.  
However, it is also worth acknowledging that this method is also able to correctly deduce the near-field properties of this structure.  To demonstrate this, in Fig.~\ref{fig:comparison}, we plot the distribution of the average (root mean square) electric field amplitude at the Fano resonance frequency, calculated from both CST and CEMDA.  
The CEMDA is inherently not able to produce the field distribution inside the nanoparticles, because the electric and magnetic fields diverge as they approach a point source.  
However, outside of the nanoparticles, the CEMDA is able to provide a near-quantitative match to the full CST simulation.  

\end{document}